\documentstyle[twoside,fleqn,espcrc2]{article}

\newcommand{\lesssim}{ {\
\lower-1.2pt\vbox{\hbox{\rlap{$<$}\lower5pt\vbox{\hbox{$\sim$}}}}\ } }
\newcommand{\gtrsim}{ {\
\lower-1.2pt\vbox{\hbox{\rlap{$>$}\lower5pt\vbox{\hbox{$\sim$}}}}\ } }

\newcommand{\be}{\begin{equation}}
\newcommand{\ee}{\end{equation}}
\newcommand{\bea}{\begin{eqnarray}}
\newcommand{\eea}{\end{eqnarray}}

\newcommand{\nn}{\nonumber}

\newcommand{\cO}{{\cal O}}
\newcommand{\cH}{{\cal H}}

\newcommand{\Imm}{\mbox{\rm Im}}

\newcommand{\MeV}{\mbox{\rm MeV}}

\newcommand{\foor}{\mbox{\rm for}}

\newcommand{\Lac}{\Lambda_{\chi}}
\newcommand{\Lmsb}{\Lambda_{\overline{\mbox{\rm\footnotesize MS}}}}

\newcommand{\exxp}{\mbox{\rm exp.}}

\newcommand{\QCD}{QCD$(\infty)\;$}

\newcommand{\N}{N_c}

\newcommand{\EM}{\mbox{\tiny\rm  EM}}

\newcommand{\ra}{\rightarrow}

\newcommand{\AmS}{{\protect\the\textfont2
  A\kern-.1667em\lower.5ex\hbox{M}\kern-.125emS}}

\title{Large--$N_c$ QCD and Spontaneous Chiral Symmetry Breaking}

\author{Eduardo de Rafael\address{Centre de Physique Th\'eorique,   CNRS-Luminy,
Case 907, F-13288 Marseille Cedex 9, France}
        \thanks{Invited talk given at the conference ``QCD 98", Montpellier,
France, July 1998. This work has been supported in part by TMR, EC-Contract
No. ERBFMRX-CT980169 (EURODA$\phi$NE).}}

\begin{document}

\begin{abstract}
I report on recent work done in collaboration with Marc Knecht~\cite{KdeR97} on
patterns of spontaneous chiral symmetry breaking in the large--$\N$ limit of
QCD--like theories,  and with Santi Peris and Michel Perrottet~\cite{PPdeR98}
concerning the question of matching long and short distances in large--$N_c$
QCD.
\end{abstract}

\maketitle

\section{INTRODUCTION}

The study of QCD in the limit of a large number of colours $N_c$ was
suggested by 't Hooft~\cite{'tH74}, soon after the discovery
of asymptotic freedom, as a way to get an insight into non--perturbative
properties of the theory. In spite of the efforts of many good
theorists~\footnote{See e.g. refs.~\cite{'tH74,RV77,W79,CW80,Ma98}}, QCD in the
large--$N_c$ limit remains an unsolved problem as yet; though the research in
this field has given rise to a series of remarkable theoretical developments
like matrix models and two--dimensional quantum gravity, and is now coming back
to QCD, as discussed by David Gross in this meeting, with promising
perspectives.

It has been shown by Coleman and Witten~\cite{CW80} (see also
refs.\cite{VaWi84,'tH80,FSBY81,CG82},) that if QCD with
$N_c=3$ confines, and if confinement persists in the large--$N_c$ limit then, in
this limit, the chiral
$U(n_f)\times U(n_f)$ invariance of the Lagrangian with $n_f$ flavours of
massless quarks is spontaneously broken down to the diagonal $U(n_f)$ subgroup. 
The hadronic spectrum of QCD in the large--$N_c$ limit, which we shall denote
\QCD for short, consists then of an infinite number of narrow states with
specific quantum numbers. In this talk I shall discuss, within the framework of
QCD($\infty$), how the ordering of states in the spectrum is related to the size
of the local order parameters of spontaneous chiral symmetry breaking
(S$\chi$SB). I shall show that the local order parameters of S$\chi$SB which
govern the operator product expansion (OPE)  and the non--local order parameters
of S$\chi$SB which govern the couplings of the low--energy effective Lagrangian
of chiral perturbation theory obey duality properties. Our discussion
will focus on the left--right correlation function
$\Pi_{LR}(Q^2)$, (see eqs.~(\ref{eq:lrtpf}) to (\ref{eq:lritpf}) below,) but the
properties I discuss for this function are rather common to any
correlation function which is an order parameter, though the details have to be
discussed separately for each Green's function.        

\subsection{The Left--Right Correlation Function}

The correlation function
$\Pi_{LR}(Q^2)$ is the invariant amplitude of the two--point function
($Q^2\equiv -q^2\ge 0$ for $q^2$ space--like)
\bea\label{eq:lrtpf}
\lefteqn{\Pi_{LR}^{\mu\nu}(q)=} \nn \\
 & & 2i\int d^4 x\,e^{iq\cdot x}\langle 0\vert
\mbox{\rm T}\left(L^{\mu}(x)R^{\nu}(0)^{\dagger}
\right)\vert 0\rangle\,,
\eea 
with currents
\be
R^{\mu}\left(L^{\mu}\right)=
\bar{d}(x)\gamma^{\mu}\frac{1}{2}(1\pm\gamma_{5})u(x)\,.
\ee 
In the chiral limit
\be\label{eq:lritpf}
\Pi_{LR}^{\mu\nu}(Q^2)=(q^{\mu}q^{\nu}-g^{\mu\nu}q^2)\Pi_{LR}(Q^2)\,.
\ee

The function $\Pi_{LR}(Q^2)$  vanishes order by order in 
perturbation theory and
is an order parameter of S$\chi$SB for all values of $Q^2$. 
It also governs the
electromagnetic $\pi^{+}-\pi^{0}$ mass difference~\cite{Lowetal67}
\be\label{eq:piem}
m_{\pi^+}^{2}\vert_{\EM}=\frac{\alpha}{\pi}\,\frac{3}{4f_{\pi}^2}\,
\int_0^\infty dQ^2\,\left(-Q^2\Pi_{LR}(Q^2)\right)\,.
\ee 
This integral converges in the ultraviolet region because~\cite{SVZ79}
\bea \label{eq:OPE}
\lefteqn{\lim_{Q^2\ra\infty} Q^6\Pi_{LR}(Q^2)=}Ê\\
 & & \left(-4\pi^2\frac{\alpha_s}{\pi}+\cO(\alpha_s^2)\right)
\langle\bar{\psi}\psi\rangle^2\,.
\eea
This behaviour also entails the two Weinberg sum rules~\cite{We67} 
in the chiral limit. 
Witten~\cite{W83} has furthermore shown that, under rather general assumptions
which are less restrictive than the large--$\N$ limit,
\be \label{eq:witten} 
-Q^2\Pi_{LR}(Q^2)\ge 0\quad\foor\quad 0\le Q^2\le\infty\,,
\ee 
which in particular ensures the positivity of the integral in
eq.~(\ref{eq:piem}) and thus the stability of the QCD vacuum with respect to 
small perturbations induced by electromagnetic interactions.
The same two--point function 
$\Pi_{LR}(Q^2)$
governs the full electroweak $\pi^{+}-\pi^{0}$ mass difference at the
one--loop level in the electroweak interactions of the Standard Model and to
lowest order in the chiral expansion~\cite{KPdeR98}.

The low $Q^2$ behaviour of this self--energy function is governed by
chiral perturbation theory:
\be 
-Q^2\Pi_{LR}(Q^2)=f_{\pi}^2+4L_{10}Q^2+\cO(Q^4)\,,
\ee 
where $L_{10}$ is one of the Gasser--Leutwyler coupling constants~\cite{GL85} of
the
$\cO(p^4)$ low energy effective chiral Lagrangian, i.e. the Lagrangian
formulated in terms of Goldstone degrees of freedom and external local 
sources only.

In \QCD the spectral function associated with
$\Pi_{LR}(Q^2)$ consists of the difference of an infinite number of narrow
vector states and an infinite number of narrow axial--vector states, together
with the Goldstone pole of the pion:
\bea
\lefteqn{\frac{1}{\pi}\Imm\Pi_{LR}(t) =   
\sum_{V}f_{V}^2 M_{V}^2\delta(t-M_{V}^2)} \nn \\
 & & -f_{\pi}^2\delta(t) -\sum_{A}f_{A}^2
M_{A}^2\delta(t-M_{A}^2)\,.
\eea 
Since $\Pi_{LR}(Q^2)$ obeys an unsubtracted dispersion relation, it follows
that
\bea\label{eq:LRN1} 
\lefteqn{-Q^2\Pi_{LR}(Q^2)=f_{\pi}^2 + \sum_{A}f_{A}^2
M_{A}^2\frac{Q^2}{M_{A}^2+Q^2}}Ê\nn \\
 & & -\sum_{V}f_{V}^2 M_{V}^2\frac{Q^2}{M_{V}^2+Q^2}\,.
\eea
Furthermore, the two Weinberg sum rules that follow from eq.~(\ref{eq:OPE})
constrain the couplings and masses of the narrow states as follows:
\be\label{eq:weinbergsr1}
\sum_{V}f_{V}^2 M_{V}^2-\sum_{A}f_{A}^2 M_{A}^2=f_{\pi}^2
\ee
and
\be\label{eq:weinbergsr2}
\sum_{V}f_{V}^2 M_{V}^4-\sum_{A}f_{A}^2 M_{A}^4=0\,,
\ee 
ensuring the convergence of the integral in eq.~(\ref{eq:piem}) in
QCD($\infty$).

\section{SPECTRAL CONSTRAINTS}

In \QCD there exists an infinite number of Weinberg--like sum rules
associated with the $\Pi_{LR}(Q^2)$--function. With the two constraints in
eqs.~(\ref{eq:weinbergsr1}) and (\ref{eq:weinbergsr2}) incorporated in the
r.h.s. of eq.~(\ref{eq:LRN1}), the large--$Q^2$ expansion of $\Pi_{LR}(Q^2)$
becomes
\bea\label{eq:largeQ}
\lefteqn{
\Pi_{LR}(Q^2) = 
\left(\sum_{V} f_{V}^2 M_{V}^6 - \sum_{A} f_{A}^2
M_{A}^6\right)\frac{1}{Q^6}}Ê\nn
\\ &  & + \left(\sum_{V} f_{V}^2 M_{V}^8 - \sum_{A} f_{A}^2
M_{A}^8\right)\frac{1}{Q^8} +
\cdots \,.
\eea
Matching this expansion in powers of $1/Q^2$ with the corresponding OPE of
$\Pi_{LR}(Q^2)$ in
\QCD leads to relations between hadronic parameters and the local order
parameters of S$\chi$SB which appear as vacuum expectation values of composite
operators in the OPE. For example, matching the $1/Q^6$--coefficient in
eq.~(\ref{eq:largeQ}) with the result in eq.~(\ref{eq:OPE}), we have that
\bea
\lefteqn{
\sum_{V} f_{V}^2 M_{V}^6 - \sum_{A} f_{A}^2
M_{A}^6=}Ê\nn \\
 & & \left(-4\pi^2\frac{\alpha_s}{\pi}+\cO(\alpha_s^2)\right)
\langle\bar{\psi}\psi\rangle^2\,.
\eea
[Notice that the negative sign in the r.h.s. above is certainly in accordance
with Witten's positivity constraint in eq.~(\ref{eq:witten}).] In full
generality, positive  moments of the
$\Pi_{LR}$--spectral function correspond to local order
parameters of S$\chi$SB $\langle
\Phi^{2n}\rangle$  of  dimension $2n$, $n=3,4,\dots$
\bea
\lefteqn{
\int_{0}^{\infty} dt t^{n-1}\left[\frac{1}{\pi}\Imm\Pi_{V}(t)-
\frac{1}{\pi}\Imm\Pi_{A}(t)\right]=} \nn \\
 & & \sum_{V} f_{V}^2 M_{V}^{2n} - \sum_{A} f_{A}^2 M_{A}^{2n}=C_{2n}\langle
\Phi^{2n}\rangle\,,
\eea
with $C_{2n}$ the corresponding short--distance Wilson coefficient calculable
in perturbative \QCD (p\QCD).

On the other hand, inverse moments of the spectral function associated with
$\Pi_{LR}$, with the pion pole removed, (this is the meaning of the tilde
symbol on top of
$\Imm\tilde{\Pi}_{A}(t)$ in eq.~(\ref{eq:invmom}) below,) determine the couplings
of the low--energy effective chiral Lagrangian. For example,
\bea\label{eq:invmom}
\lefteqn{
\int_{0}^{\infty} dt \frac{1}{t}\left[\frac{1}{\pi}\Imm\Pi_{V}(t)-
\frac{1}{\pi}\Imm\tilde{\Pi}_{A}(t)\right]=} \nn \\
 & & \sum_{V} f_{V}^2 -\sum_{A} f_{A}^2 = -4 L_{10}\,.
\eea
Moments with higher inverse powers of $t$ are associated with couplings of
composite operators of higher dimension in the chiral Lagrangian. 

\subsection{Patterns of S$\chi$SB}

In ref.~\cite{KdeR97} we have shown how the ordering of vector and axial
vector states in the hadronic spectrum of \QCD is correlated to the size of
the local order parameters $\langle
\Phi^{2n}\rangle$ of S$\chi$SB. Quite generally, we have shown that 
in \QCD {\it spontaneous chiral
symmetry breaking \`{a} la Nambu--Goldstone with
$f_{\pi}^2\not=0$ necessarily implies the existence of non--zero local order
parameters which transform according to the representation
$(n_f,\bar{n}_f)+(\bar{n}_f,n_f)$ of the chiral group.} This is in a way the
converse theorem~\footnote{This converse theorem, however, was already known to
{\it all the experts} according to J.~Stern.} to the Coleman--Witten
theorem~\cite{CW80} stated in the Introduction.

The minimal pattern of a spectrum compatible with the short--distance properties
of the
$\Pi_{LR}$--function in \QCD with $n_{f}=3$, is
one which besides the Goldstone pseudoscalar nonet has a vector nonet
of states and an axial--vector nonet of states. The
required ordering is then
$M_{V}< M_{A}$. Implicit here, of course, is the assumption that the sum of the
infinite number of narrow vector states and the sum of the infinite number of
narrow axial--vector states with masses higher than the highest mass explicitly
considered (here
$M_{A}$) are already dual to their respective p\QCD continuum. Their
contributions to the
spectral function
$\Imm\Pi_{LR}(t)$ cancel then each other and, therefore, they vanish in the
Weinberg sum rules, as well as in the generalized Weinberg sum rules
discussed above. This minimal pattern is also the one that the authors of
ref.~\cite{Lowetal67} considered in their evaluation of the electromagnetic pion
mass difference, which gives as a result:
$\Delta m_{\pi}=5.2\,\MeV$ remarkably close to the experimental result:
$\Delta m_{\pi}\vert_{\exxp}=4.59\,\MeV$. As shown in ref.~\cite{KdeR97}
the extreme version of the so called {\it generalized}
$\chi$PT proposed by J.~Stern {\it et al.}~\cite{FSSKM}, where
$\langle\bar{\psi}\psi\rangle=0$, is incompatible with this phenomenologically
successful minimal pattern of a hadronic low--energy spectrum with only one
$V$--state and only one $A$--state.

Another interesting feature discussed in ref.~\cite{KdeR97} 
is the possible existence of low--energy particle spectra in vector--like  gauge
theories with a rather different structure than the one observed in the QCD
hadronic spectrum. For example, the minimal pattern required to have a negative
electroweak
$S$ parameter (the equivalent of $-4 L_{10}$ in an underlying technicolour--like
model of electroweak breaking,) is a spectrum with two axial--vector states
$A_{1}$ and $A_{3}$ and a vector state $V_{2}$ with an increasing ordering of
masses: $M(A_{1})<M(V_{2})<M(A_{3})$.

\subsection{Duality Properties of $\Pi_{LR}(Q^2)$}

It is instructive to reconsider the two--point function $\Pi_{LR}(Q^2)$
in the simple case of a minimal spectrum with one vector
state $V$ and one axial--vector state $A$. In this case $\Pi_{LR}(Q^2)$ reduces
to a very simple form
\bea
\lefteqn{ 
-Q^2
\Pi_{LR}(Q^2) = f_{\pi}^2\frac{1}{\left(1+\frac{Q^2}{M_{V}^2}\right)
\left(1+\frac{Q^2}{M_{A}^2}\right)}} \nn \\
 &  &  =f_{\pi}^2 \frac{M_{A}^2
M_{V}^2}{Q^4}\frac{1}{\left(1+\frac{M_{V}^2}{Q^2}\right)
\left(1+\frac{M_{A}^2}{Q^2}\right)}\,.
\eea 
This equation shows explicitly a remarkable short--distance
$\rightleftharpoons$ long--distance symmetry. Indeed, with $g_{A}$ defined so
that $M_{V}^2=g_{A}M_{A}^2$ and
$z\equiv\frac{Q^2}{M_{V}^2}$, then
\be 
-Q^2\Pi_{LR}(Q^2)\equiv f_{\pi}^2\cH(z;g_{A})\,,
\ee 
and we find that
\be
\cH(z;g_{A})=\frac{1}{z^2}\frac{1}{g_{A}}
\cH\left(\frac{1}{z};\frac{1}{g_{A}}\right)\,.
\ee 
This means that, in the minimal pattern spectrum, the {\it non--local order
parameters} corresponding to the long--distance expansion for $z\ra 0$, which
correspond to couplings of the effective chiral Lagrangian i.e., 
\bea
\lefteqn{ 
-Q^2\Pi_{LR}(Q^2)=
f_{\pi}^2\left\{1-(1+g_{A})z\right.}Ê\nn \\
 & & \left. +(1+g_{A}+g_{A}^2)z^2+\cdots\right\}\,,
\eea 
are exactly correlated to the {\it local order parameters} of the
short--distance OPE for $z\ra\infty$ in a very simple
way:
\bea
\lefteqn{-Q^2
\Pi_{LR}(Q^2)= f_{\pi}^2\frac{1}{g_{A}}\frac{1}{z^2}\times
\left\{1- \left(1+\frac{1}{g_{A}}\right)\frac{1}{z}\right.} \nn \\
& & \left. +
\left(1+\frac{1}{g_{A}}+\frac{1}{g_{A}^2}\right)\frac{1}{z^2}+\cdots\right\}\,;
\eea
in other words, knowing the expansion at large $z$ we can reconstruct the
corresponding expansion at small $z$ and vice versa.

\section{APPROXIMATED \QCD}

The minimal pattern of an acceptable hadronic spectrum in \QCD turns out to be
rather successful in describing global features of low--energy hadron
phenomenology. Partly inspired by the traditional successes of ``vector meson
dominance'' in predicting, e.g., the low--energy constants of the effective
chiral Lagrangian~\cite{EGPR89,EGLPR89} we have recently proposed~\cite{PPdeR98}
to consider  the approximation to \QCD which restricts the hadronic spectrum in
the channels with
$J^P$ quantum numbers $0^{-}$, $1^{-}$, $0^{+}$ and
$1^{+}$  to the lowest energy state and treats the rest of the narrow states as a
p\QCD continuum, the onset of the continuum being fixed by
consistency constraints from the operator product expansion; (like the absence
of
$d=2$ operators.) We have shown
that there exists a useful effective Lagrangian description of this well defined
lowest meson dominance (LMD) approximation to
QCD($\infty$). The degrees of freedom in the effective Lagrangian are then a
nonet of pseudoscalar Goldstone particles which are collected in a unitary
matrix $U(x)$, and nonets of vector fields $V(x)$, scalar fields $S(x)$ and
axial--vector fields
$A(x)$ associated with the lowest energy states of the hadronic spectrum which
are retained. We have derived the effective Lagrangian by implementing
successive requirements on an extended Nambu--Jona-Lasinio (ENJL)--type
Lagrangian~\cite{NJL61,DSW85,BBdeR93,BdeRZ94,Bi96} which we have chosen as the
initial {\it ansatz}. The first requirement is to eliminate the effects of {\it
non--confining}
$Q\bar{Q}$ discontinuities ($Q$ denotes the constituent quark field) by
introducing an infinite number of appropriate local operators with couplings
which can be fixed in terms of the three parameters of the starting
ENJL--Lagrangian itself, i.e. the coupling constants
$G_S$,
$G_V$ and the scale $\Lac$.  We have shown that the {\it matching} of the
two--point functions of this effective Lagrangian to their \QCD short--distance
behaviour can be systematically implemented. In particular, the first and second
Weinberg sum rules are automatically satisfied.
 
For Green's functions beyond two--point functions, the removal of the {\it
non--confining}
$Q\bar{Q}$ discontinuities produced by the initial ENJL {\it ansatz} is however
not enough to guarantee in general the correct {\it matching} to the leading QCD
short--distance behaviour and further local operators have to be included. We
have discussed this explicitly in the case of the VPP and VPA three--point
functions, and shown that the {\it matching} with the QCD short--distance
leading behaviour which follows from the OPE restricts the initial three free
parameters of the ENJL--Lagrangian {\it ansatz} to just one mass scale $M_{Q}^2$
and a dimensionless constant
$M_{Q}^2/\Lac^2$. The resulting low--energy Lagrangian in the vector and
axial--vector sector, and  to
$\cO(p^4)$ in the chiral expansion, coincides with the class of phenomenological
Lagrangians discussed in ref.~\cite{EGLPR89} which also have two free parameters
$f_{\pi}^2$ and $f_{\pi}^{2}/M_{\rho}^{2}$. In this respect, this explains the
relation to QCD of the phenomenological VMD Lagrangians discussed in
ref.~\cite{EGLPR89}: to
$\cO(p^4)$, they can be viewed as the effective low--energy Lagrangians of the
LMD approximation to QCD($\infty$). On the other hand, the fact that the
resulting low--energy Lagrangian coincides with the phenomenological VMD
Lagrangians discussed in ref.~\cite{EGLPR89} demystifies to a large extent the
r\^{o}le of the ENJL--Lagrangian itself  as a fundamental step in deriving the
low--energy effective Lagrangian of QCD. The ENJL--Lagrangian turns out to be
already a very good {\it ansatz} to describe in terms of quark fields degrees of
freedom the LMD approximation to
\QCD and this is why it is already quite successful at the phenomenological
level; but, when the {\it non--confining} $Q\bar{Q}$
discontinuities are systematically removed, the phenomenological predictions
improve even more.

There is an advantage, however, in starting with the ENJL--Lagrangian as an {\it
ansatz}, and that is that in this description of the LMD approximation to
QCD($\infty$), all the couplings to all orders in
$\chi$PT are clearly correlated to the same two free parameters. For example the
$L_5$ and $L_8$ constants, as well as part of the contribution to the $L_3$
constant which result from scalar exchanges, are now proportional to a
universal dimensionless parameter, while in a purely phenomenological description
in terms of chiral effective Lagrangians which include resonances as discussed
e.g. in refs.~\cite{EGPR89} and references therein, these
low--energy constants require new phenomenological input.   
   
Finally, we wish to insist on the phenomenological successes of the LMD
approximation to
\QCD as demonstrated by the results presented in ref.~\cite{PPdeR98}. In
particular the Gasser--Leutwyler coupling constants of the $\cO(p^4)$
chiral Lagrangian which do not trivially vanish in the large--$\N$ limit, are all
fixed by the ratio $f_{\pi}^2/M_{V}^2$:
\bea
\lefteqn{
 6L_1 =3L_2=-\frac{8}{7}L_3 = 4 L_5 = 8L_8}Ê\nn \\
& &  =\frac{3}{4} L_9
=-L_{10}=\frac{3}{8}\frac{f_{\pi}^2}{M_{V}^2}\,.
\eea  
These results show that the LMD approximation to \QCD is indeed a very good
approximation to full fledged QCD. The deep reason for that may very well be
correlated to the size of $\Lmsb^{(3)}$. Indeed, the onset of the pQCD
continuum associated with the LMD approximation in the case, e.g., of the vector
two--point function is at a $t$ value~\cite{PPdeR98}: $t_0\simeq 1.5\,GeV^2$,
which is already sufficiently high for p\QCD to be applicable if
$\Lmsb^{(3)}\simeq 400\,\MeV$ as known phenomenologically.    

It seems now worthwhile to apply the LMD approximation to \QCD to the calculation
of couplings of
$\cO(p^6)$ and $\cO(e^2 p^2)$ as well. One can also, at last, reconsider
non--leptonic weak interactions in the light of this effective Lagrangian
framework with some hope of success, since within this approach we expect to be
able to show a rather good {\it matching} between the long--distance evaluation
of matrix elements of four--quark operators and the short--distance pQCD
logarithmic dependence of the Wilson coefficients~\footnote{A first successful
example of this program has already appeared in ref.~\cite{KPdeR98}.} .

\section{ACKNOWLEDGEMENTS}

It is a pleasure to thank my colleagues Marc Knecht, Santi Peris and Michel
Perrottet, with whom the work reported here has been done, for a very pleasant
collaboration.


\begin{thebibliography}{99}

\bibitem{KdeR97}
         M.~Knecht and E.~de Rafael, Phys. Lett. {\bf B424} (1998)
         335.

\bibitem{PPdeR98}
         S.~Peris, M.~Perrottet and E.~de Rafael, JHEP {\bf 05}
         (1998) 011.

\bibitem{'tH74}
        G.~'t Hooft, Nucl. Phys. {\bf B72} (1974) 461; {\it ibid.} Nucl. Phys.
        {\bf B73} (1974) 461.

\bibitem{RV77}
        G.~Rossi and G.~Veneziano, Nucl. Phys. {\bf B123} (1977) 507.

\bibitem{W79}
        E.~Witten, Nucl. Phys. {\bf B160} (1979) 57.

\bibitem{CW80}
	       S.~Coleman and E.~Witten, Phys. Rev. Lett. {\bf 45} (1980) 100.

\bibitem{Ma98}
        A.~Manohar, ``Large N QCD'', lectures at the Les Houches
        Summer School 1997, hep-ph/9802419.

\bibitem{VaWi84}
        C.~Vafa and E.~Witten, Nucl. Phys. {\bf B234} (1984) 173.

\bibitem{'tH80}
        G.~'t Hooft, in Recent Developments in Gauge theories, G.~'t Hooft et
        al., eds. (Plenum, New York, 1980).

\bibitem{FSBY81}
        Y.~Frishman, A.~Schwimmer, T.~Banks and S.~Yankielowicz, Nucl. Phys.
        {\bf B177} (1981) 117.

\bibitem{CG82}
        S.~Coleman and B.~Grossman, Nucl. Phys. {\bf B203} (1982) 205.

\bibitem{Lowetal67}
        T.~Das, G.S~Guralnik, V.S.~Mathur, F.E.~Low and J.E.~Young, Phys.
        Rev. Lett. {\bf 18} (1967) 759.

\bibitem{SVZ79}
        M.A.~Shifman, A.I.~Vainshtein and V.I.~Zakharov, Nucl. Phys. {\bf
        B147} (1979) 385, 447.

\bibitem{We67}
        S.~Weinberg, Phys. Rev. Lett. {\bf 18} (1967) 507.

\bibitem{W83}
        E.~Witten, Phys. Rev. Lett. {\bf 51} (1983) 2351.

\bibitem{KPdeR98}
		      M.~Knecht, S.~Peris and E.~de Rafael, hep-ph/9809594.

\bibitem{GL85}
        J.~Gasser and H.~Leutwyler, Nucl. Phys. {\bf B250} (1985) 465.

\bibitem{FSSKM}
        N.H.~Fuchs, H.~Sazdjian and J.~Stern, Phys. Lett. {\bf B269} (1991)
        183;\\
        J.~Stern, H.~Sazdjian and N.H.~ Fuchs, Phys. Rev. {\bf D47} (1993)
        3814;\\
        M.~Knecht, B.~Moussallam, J.~Stern and N.H.~Fuchs, Nucl. Phys. {\bf B457}
        (1995) 513; and references therein.

\bibitem{EGPR89}
         G.~Ecker, J.~Gasser, A.~Pich and E.~de Rafael, Nucl. Phys. {\bf B321}
         (1989) 311.

\bibitem{EGLPR89}
         G.~Ecker, J.~Gasser, H.~Leutwyler, A.~Pich and E.~de Rafael, Phys.
         Lett. {\bf B223} (1989) 425.


\bibitem{NJL61}
         Y.~Nambu and G.~Jona-Lasinio, Phys. Rev. {\bf 122} (1961) 345.

\bibitem{DSW85}
         A.~Dahr, R.~Shankar and S.~Wadia, Phys. Rev. {\bf D31} (1985)
         3256.

\bibitem{BBdeR93}
         J.~Bijnens, C.~Bruno and E.~de Rafael, Nucl. Phys. {\bf B390}
         (1993) 501.

\bibitem{BdeRZ94}
         J.~Bijnens, E.~de Rafael and H.~Zheng, Z. Phys.
        {\bf C62} (1994) 437.

\bibitem{Bi96}
         J.~Bijnens, Phys. Rep. {\bf 265} (1996) 369.

\end{thebibliography}
\end{document}